\documentclass[journal]{IEEEtran}
\usepackage{blindtext}
\usepackage{graphicx}
\usepackage{cite}
\usepackage{hyperref}
\usepackage{authblk}
\usepackage{color}
\usepackage{caption}
\usepackage{tabularx}
\usepackage{verbatim}
\usepackage{caption}
\usepackage{amsmath}
\usepackage{fixltx2e}
\usepackage{array}
\usepackage{multirow}
\usepackage{url}
\IEEEoverridecommandlockouts

\usepackage{fancyhdr}

\usepackage[numbers,sort&compress]{natbib}

\usepackage{algorithm}  
\usepackage{algpseudocode}  
\usepackage{amsmath}  

\newcommand{\tabincell}[2]{\begin{tabular}{@{}#1@{}}#2\end{tabular}}

\begin{document}
\title{Mobile-end Tone Mapping based on Integral Image and Integral Histogram}

\author[2]{Jie Yang  \thanks{Corresponding author e-mail: yangjie@westlake.edu.cn}}
\author[1]{Mengchen~Lin} 
\author[1]{Ziyi Liu} 
\author[1]{Ulian Shahnovich}
\author[1]{Orly Yadid-Pecht}

\affil[1]{I2Sense lab, University of Calgary, Calgary T2N 1N4, Canada}
\affil[2]{Westlake University, Hangzhou  310024, China}

\maketitle

\thispagestyle{fancy}
\fancyhead{}
\lhead{}
\cfoot{}
\rfoot{}
	
\begin{abstract}
Wide dynamic range (WDR) image tone mapping is in high demand in many applications like film production, security monitoring, and photography. It is especially crucial for mobile devices because most of the images taken today are from mobile phones, hence such technology is highly demanded in the consumer market of mobile devices and is essential for good customer experience. However, high quality and high-performance WDR image tone mapping implementations are rarely found in the mobile-end.
In this paper, we introduce a high performance, mobile-end WDR image tone mapping implementation. It leverages the tone mapping results of multiple receptive fields and calculates a suitable value for each pixel. The utilization of integral image and integral histogram significantly reduce the required computation. Moreover, GPU parallel computation is used to increase the processing speed. The experimental results indicate that our implementation can process a high-resolution WDR image within a second on mobile devices and produce appealing image quality.
\end{abstract}

\begin{IEEEkeywords}
Wide dynamic range, compression, mobile graphic processing unit (GPU), tone mapping, parallel computing.
\end{IEEEkeywords}

\section{Introduction}

\IEEEPARstart{W}{ide} dynamic range is defined as the ratio of the intensity of the brightest point to that of the darkest point in a scene or image. Traditional display devices such as LCD, CRT, and LED are usually limited to 8 bits Thus they can only represent the dynamic range of 255:1. However, the fast developing image sensor technology and image processing algorithms enable the capture of WDR images with a much wider dynamic range than that of the display devices. Therefore it is impossible to properly reproduce the WDR image on the display directly. Tone mapping algorithms are often called tone mapping operators (TMO), they serve the goal of compressing the WDR image to match the dynamic range of the display devices. The development of tone mapping algorithms started to emerge in the early 1990s. Tumblin and Rushmeier \cite{tumblin1993tone} and Ward \cite{ward1994contrast} did the earliest attempts. Tumblin and Rushmeier aimed to match the perceived brightness of the displayed image with that of the scene. Ward used a linear scaling, focusing on preserving the image contrasts. 
These algorithms are classified as global tone mapping functions because they use the same tone-curve for all the pixels of the WDR image. 
In general, global tone mapping algorithms are computationally easy and mostly artifact-free. Hence, they have unique advantages in implementations. However, there are only a few global TMO implementations found in the literature because they are prone to loose details and contrast in either bright or dark regions due to the global compression feature. Drago \textit{et al.} \cite{drago2003adaptive} proposed a function that could change the base of the logarithmic function according to the pixel brightness. It is one of the most commonly used examples of tone mapping in various publications. The sigmoid function is similar to the response curve of the human visual system (HVS), it is thus used in many bio-inspired TMOs. Works like \cite{schlick1995quantization, van2006encoding, benoit2009spatio,reinhard2005dynamic} try to simulate the procedure of dynamic range compression of the HVS by mimicking the response curve of our photo-receptors. Although these approaches may be effective in reducing the dynamic range, they have an inherent flaw that the tone mapped image represents the internal representation rather than the luminance which is more expected by our eyes.

In many research works, tone mapping is regarded as a constrained optimization problem. The objective is to achieve a tone-mapping that is most preferred in terms of objective quality. Mantiuk \textit{et al.} tried to minimize visible distortion during the tone mapping \cite{mantiuk2008display}. Ma \textit{et al.} proposed a tone mapping method that can optimize the tone mapped image quality index \cite{ma2014high}. Recently, tone mapping with edge preserving filter has become the most popular way to tone map WDR images \cite{durand2002fast,farbman2008edge, he2010guided,gu2013local,paris2015local}. It first applies an edge-preserving filter to the log domain of the WDR image's luminance channel. The edge-preserving filter separates the image into two layers, namely a base layer that contains global brightness information and a detailed layer that mainly consists of local texture. The base layer is processed with the compressive tone-curve while the detail layer is untouched. Thus, local details could be preserved while the overall dynamic range is reduced. Finally, the base-layer and the detail-layer are combined and transformed back to the original linear domain for display. 

Most tone mapping algorithms are implemented in desktop-end. However, there are several GPU implementations made in recent years, which demonstrate great computational efficiency with the help of GPU's parallel processing ability. 
Chen \textit{et. al} achieved 50 Hz tone mapping by developing a new data structure and paralleling their edge preserving filter on GPU \cite{chen2007real}. Akyuz demonstrated a WDR imaging pipeline realized by GPU \cite{akyuz2015high}. It yields 2 to 3 orders of performance improvement when compared to the CPU implementation. Urena \textit{et. al} evaluated both GPU and FPGA performances on a new tone mapping algorithm \cite{urena2012real}. When compared with CPU implementation, speed-up factors of 7.5 and 15 are achieved for the GPU and FPGA, respectively.

In this paper, we propose an algorithm that is inspired by the HVS, it tone maps a pixel by taking into account multiple receptive fields that surround the pixel. The utilization of integral image and integral histogram make the whole tone mapping process highly parallel. 
Experimental results including the image quality evaluation and mobile-end implementation performance are carried out to prove the effectiveness and efficiency of our work.

The rest of the paper is structured as follows. Section 2 details the proposed algorithm and section 3 describes the mobile-end implementation. Section 4 provides experimental results. The last section concludes this work.

\section{Proposed Algorithm}
When viewing WDR scenes, our HVS adopts the local adaptation mechanism to help us see the details in all parts of the scene. Local adaptation can be mainly explained as the ability to accommodate the level of a certain visual field around the current fixation point. Moreover, it also reveals that different luminance intervals could
result in overlapping reactions on the limited response range of
the visual system, thus extending our visual response range to
cope with the full dynamic range of high contrast scenes. Inspired by the concept of local adaptation, we design our tone mapping algorithm as follows.
The WDR image $i$ is first transferred to $l$ using logarithmic compression. For every pixel $l(x,y)$, one can always find $s$ different receptive fields $w_i, i =1,2...s$ where $l(x,y)$ is the center of every receptive field. 
To adapt to every visual field, we can tone map them separately. Since HVS has a logarithmic response to light intensity, we choose histogram adjustment \cite{larson1997visibility} which is one of the simplest tone mapping methods to tone map every receptive field. The processing flow of applying histogram adjustment to every receptive field $w_i$ is summarized as follow: 
first, a histogram of image luminance in the logarithmic domain is
constructed for every $w_i$. Denoting $f_{w_i}(b_i)$ as the pixel count in a bin $b_i$ of the histogram, a cumulative probability function is defined as
\begin{equation}  
P_{w_i}(b) = \frac{1}{T}\sum_{b_i < b }f_{w_i}(b_i) \ \ \ T = \sum_{b_i}f_{w_i}(b_i)
\end{equation}
The tone mapped value for the center pixel of $w_i$ is calculated using the following equation:
\begin{equation}  
L_{w_i}(x,y) = min(L_d) + ((max(L_d) - min(L_d))P(l(x,y))
\end{equation}
where $min(L_d)$ and $max(L_d)$ are the minimum and maximum display luminance, $l(x,y)$ is the center pixel of receptive field $w_i$

For each receptive field $w_i$, Eq. 2 will give a tone mapped value of $L_{w_i}(x,y)$. The values of all receptive fields will be fused together using a weight function:
\begin{equation}  
L(x,y) = \frac{\sum_{i=1}^s W_{w_i}(x,y)L_{w_i}(x,y)}{\sum_{i=1}^s W_{w_i}(x,y)} 
\end{equation}
$W_{w_i}(x, y)$ is the weight for value $L_{w_i}(x,y)$, and it is calculated using the following formula
\begin{equation}
W_{w_i}(x,y) = \frac{\sigma_{{w_i},(x,y)}^2}{\sigma_{{w_i},(x,y)}^2 + \epsilon} 
\label{m-e representation}
\end{equation}
$\sigma_{{w_i},(x,y)}^2$ is the local variance of receptive field $w_i$ centered at pixel location $(x, y)$, and $\epsilon$ is a user defined parameter. 

The proposed algorithm will be computationally expensive because for every pixel location $(x,y)$, one needs to compute $s$ different histograms and $s$ different variances. In the following section, we will show how the histograms and variances are calculated in a single pass fashion.

\subsection{Integral Image and Integral Histogram based Acceleration}
The integral image is prominently used within the Viola-Jones object detection framework from 2001 \cite{viola2004robust}. It can significantly reduce the computation burden when an accumulative sum of image area is required. The following equation defines an integral image.
\begin{equation}
 I(x,y)= \sum_{\substack{x' \leq x,\ y' \leq y}} i(x',y')
\end{equation}
where $i(x,y)$ is the value of the pixel at $(x, y)$. A great feature of integral image $I$ is that summation of any rectangular region in the original image $i$ can be computed efficiently in a single pass. For example, if there are four points $A (x_0, y_0)$, $B (x_1, y_0)$, $C (x_1, y_1)$ and $D (x_0, y_1)$ $(x_0 < x_1$, $y_0 < y_1)$ in image $i$, the accumulative summation of rectangular $w$ that is enclosed by the four points is equal to:
\begin{equation}
 \sum_{\substack{x_0<x \leq x_1,\\y_0<y \leq y_1}} i(x, y) = I(A) + I(C) - I(B) - I(D) 
\end{equation}
It can be seen from Eq. 6 that with the help of integral image, only one addition and two subtractions are required for calculating an accumulative sum.

A natural extension of integral image is the integral histogram. It is a fast way to extract histogram of a specific area \cite{porikli2005integral}. If the pixel values of an image $i$ are equally segmented into $n$ different bins of a histogram, then we can build an $n$-channel map $h$ where the $k$-th channel $h_k$ is defined as follow:
\begin{equation}
h_k (x', y')=\left\{
\begin{array}{rcl}
1      &      & {i(x, y)      \in     b_k}\\
0      &      & {i(x, y)      \notin b_k}
\end{array} \right.
\end{equation}
$b_k$ indicates the $k$-th bin. For each channel $h_k$, we can have an integral image $H_k$
\begin{equation}
H_k(x, y) = \sum_{\substack{x' \leq x, \   y'\leq y   }} h_k(x',y') 
\end{equation}
Then, we can build a $n$-channel integral histogram $H$ where the $k$-th bin of the histogram is calculated with
\begin{equation}
H_k(A) + H_k(C) - H_k(B) - H_k(D)
\end{equation}
It is evident that with the help of integral histogram. An $n$-bin histogram of any local area can be calculated easily with $n$ addition and 2$n$ subtractions.

With the help of integral histogram, the histogram calculation of Eq. 1 becomes straightforward. We first equally divide image $l$ into $n$ bins based on pixel values. $n$ is a user defined parameter. The integral histogram of the logarithmic image $l$ can be built according to Eq. 7 and Eq. 8. Then for any receptive field $w_i$, the pixel count  of $f_{w_i}(b_i)$ can be solved by Eq. 9. Hence, Eq. 1 and Eq. 2 can be derived easily. 
The local variance value $\sigma_{{w_i},(x,y)}^2$ has the following form:
\begin{equation}
\sigma_{{w_i},(x,y)}^2 = \frac{\sum_{w_i, (x,y)}i^2-(\sum_{w_i, (x,y)} i)^2}{|w_{i, (x,y)}|^2}
\end{equation}
where $\sum_{w_i, (x,y)}i^2$ and $\sum_{w_i, (x,y)} i$ are the accumulative sums of receptive field $w_i, (x, y)$ centered at pixel location $(x, y)$, and $|w_{i, (x,y)}|^2$ is the number of pixels within the receptive field. The two accumulative sums in Eq. 10 can be computed with integral image of $i$ and $i^2$.

The majority of the computational burden of Eq. 1 to Eq. 4 are optimized with the help of integral image and integral histogram. The remaining computation is just some basic operations. After the formation of the integral image and integral histogram, the entire tone mapping process becomes pixel-parallel. Which means all pixels can be computed in parallel without data dependency. This feature is significantly helpful when it comes to GPU implementation where large amounts of single-instruction-multiple-data (SIMD) processing units are available. We will show the GPU pixel-parallel implementation in Section III.

\begin{figure}[tb]
\begin{center}
    \includegraphics[scale = 0.22]{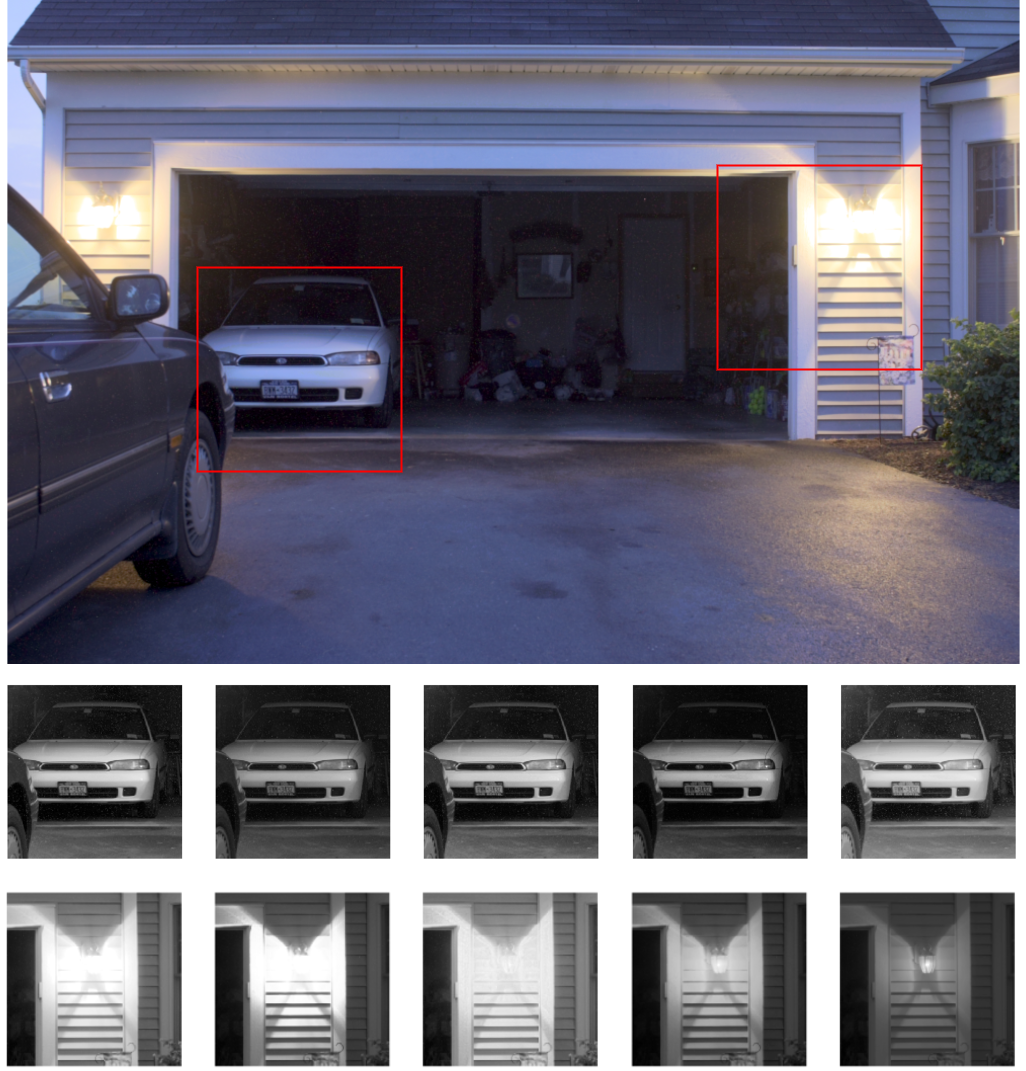}
\end{center}
\caption{Original WDR image and two same image patches from different algorithms. Top image is the original image (tone mapped for showing). From left to right of the bottom two rows are image patches taken after implementation of algorithm Mantiuk \textit{et al.} \cite{mantiuk2008display}, Durand \textit{et al.} \cite{durand2002fast}, Reinhard \textit{et al.} \cite{reinhard2002photographic}, Drago \textit{et al.} \cite{drago2003adaptive} and simple logarithmic compression.}
\label{xxx}
\end{figure}

\subsection{Analysis}
\begin{figure}[tb]
\begin{center}
   \includegraphics[scale = 0.31]{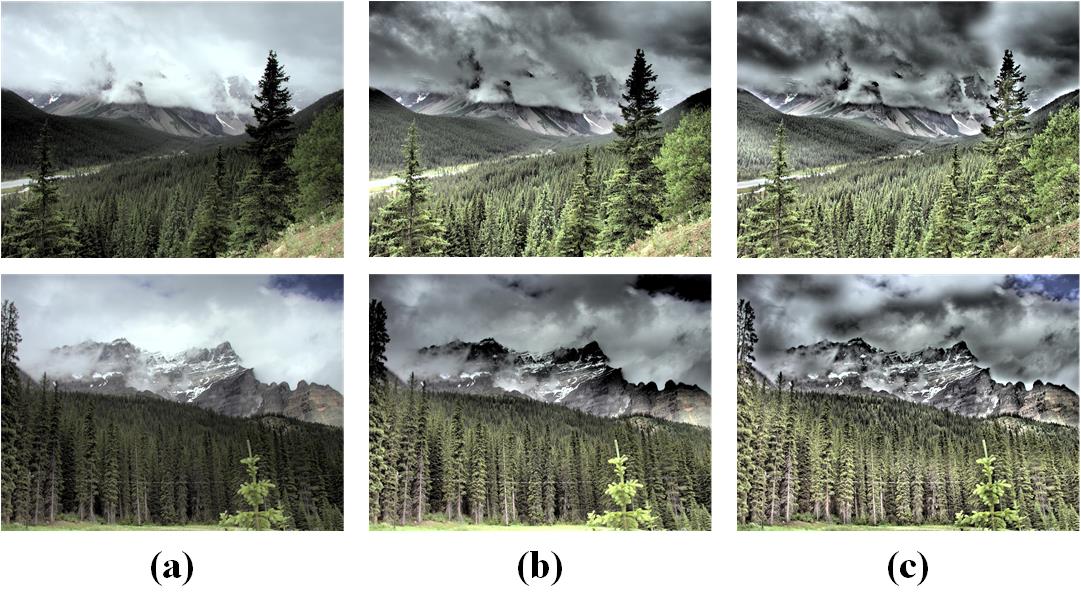}
\end{center}
   \caption{Images tone mapped with one receptive field. (a) Toned image uses receptive field which equals to image height and width. (b) Toned image uses receptive field which equals to quarter image height and width (c) Toned image uses receptive field which equals to 1/16 of image height and width.}
\label{Fig_5}
\end{figure}

In this part, we make a detailed analysis of how the proposed algorithm compresses the dynamic range while preserving local details. 

The main reason that the algorithm adopts a simple tone mapping method (Eq. 2) rather than a more complicated one to tone map every receptive field $w_i$ is because the local dynamic range of a single receptive field $w_i$ is mostly much lower than the dynamic range of the entire WDR image. Fig. 1 shows an example that a simple tone mapping method is already sufficient to produce good results. The image on top of Fig. 1 is the original WDR image. We tone map this image with different algorithms. Two patches of the same size and at the same location are selected from the corresponding tone mapped images. From left to right on the two bottom rows of Fig. 1, the image patches are from
Mantiuk \textit{et al.} \cite{mantiuk2008display}, Durand \textit{et al.} \cite{durand2002fast}, Reinhard \textit{et al.} \cite{reinhard2002photographic} and Drago \textit{et al.} \cite{drago2003adaptive}, respectively. The last column shows image patches from the original WDR image and then tone mapped with a logarithmic response function. The image patches using local processing show comparable or even better results when compared with other algorithms. In the lamp area, the image is not as saturated as the other four images, and in the car area, the image is brighter than the other four images. This example demonstrates that with local adaptation, even a simple tone mapping method can reveal the local details of wide dynamic range scenes. 

However, revealing local detail is not enough for good tone mapping, because artifacts usually emerge if there is no consideration of global statistics. Figure. 2 shows a tone mapping example of our algorithm with only one receptive field ($s = 1$ in Eq. 3). It is apparent that the smaller the receptive field, the better the details, but there are many disturbing artifacts especially at uniform areas. On the other hand, the larger the receptive field, the fewer the details as well as the artifacts. Actually, the size of the receptive field can balance between the global and local statistics during tone mapping. Consider the most extreme situation when the receptive field is as large as the WDR image itself, then the tone mapping becomes global. In this circumstance, there will be no artifacts because the tone mapping is global, it can preserve the brightness consistency of the image based on the monotonic tone mapping curve. But, the details are also concealed due to the global compression feature. If the receptive field is so small that there are only several pixels, the details will be greatly exaggerated because every receptive field will be given the maximum display range between $min(L_d)$ and $max(L_d)$ regardless of its size. If the pixels have similar values, for instance, they are all background pixels of clear sky, then, some pixel will still be tone mapped to $min(L_d)$ and become artifacts because of Eq. 2. The local compression could sabotage the image brightness consistency, especially at uniform areas because the tone mapping function of the entire WDR image is no longer monotonic. 
Although the results are significantly different when tone mapped with large and small receptive fields, it is also evident that the displayed visual appearances are complimentary which contain local detail and global brightness consistency. Since every pixel is included in $s$ different receptive fields, one solution to obtain these local and global appearances is to fuse all receptive fields using a weighted summation as shown in Eq. 3. 

To reveal local detail and maintain brightness consistency, we should consider the two possible cases of local receptive fields. In a ``flat" receptive field where pixel values are mostly the same, we should tone map it more globally to maintain its image consistency and reduce possible artifacts. In a ``high variance" receptive field where pixel values fluctuate significantly, we should tone map it more locally to reveal details. Eq. 4 is taken from guided image filter \cite{he2010guided} and it is a very efficient way to measure the ``flatten" or ``variance" degree of a local image area. If the variance value $\sigma_{{w_i},(x,y)}^2$ is much larger than $\epsilon$, the weight value will be close to 1, which indicates that the centre pixel of receptive field ${w_i},(x,y)$ is ``high variance". If the variance is much less than $\epsilon$, the weight value will be close to 0, and the center pixel is regarded as belonging to a ``flat" receptive field. Eq. 4 gives pixel-wise and receptive-field-wise weight for each pixel and it has edge preserving feature that is helpful for artifacts reduction.
Fig. 3 shows the weight map calculated with different receptive field. Bright pixels mean the computed weight value is close to one and dark pixels mean that it is close to zero. It can be seen that in smaller scales, the weight can better preserve details such as edges and textures.

\begin{figure}[tb]
\begin{center}
   \includegraphics[scale = 0.31]{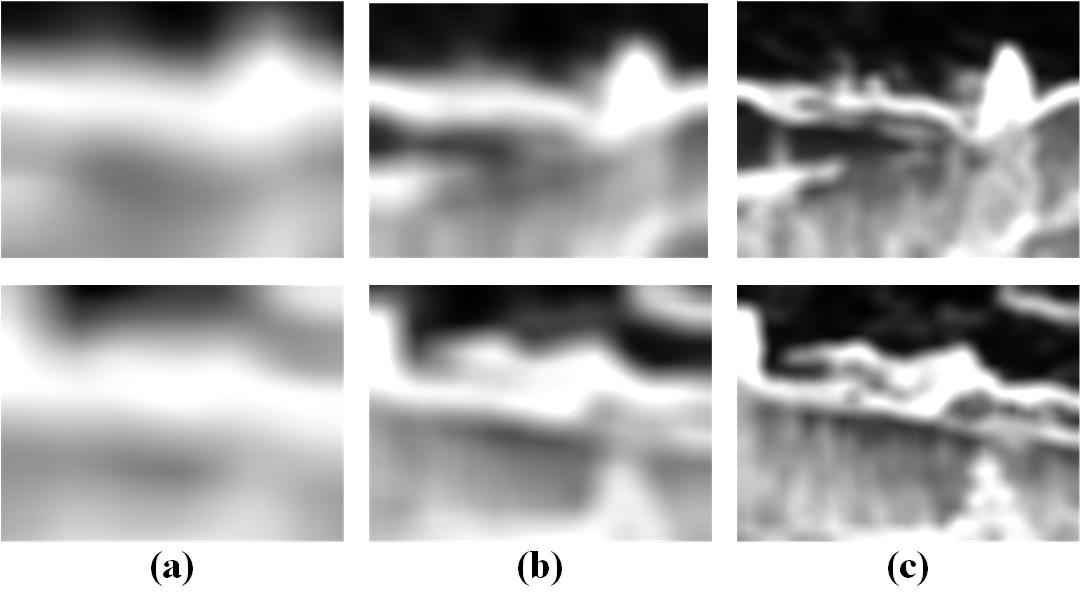}
\end{center}
   \caption{Weight maps computed using Eq. 4. (a) $w_i$ is equal to quarter image  height and width of image. (b) $w_i$ is equal to 1/8 image height and width of image. (c) $w_i$ is equal to 1/16 image height and width of image.}
\end{figure}

\subsection{Parameter Setting}
\begin{figure}[tb]
\begin{center}
   \includegraphics[scale = 0.21]{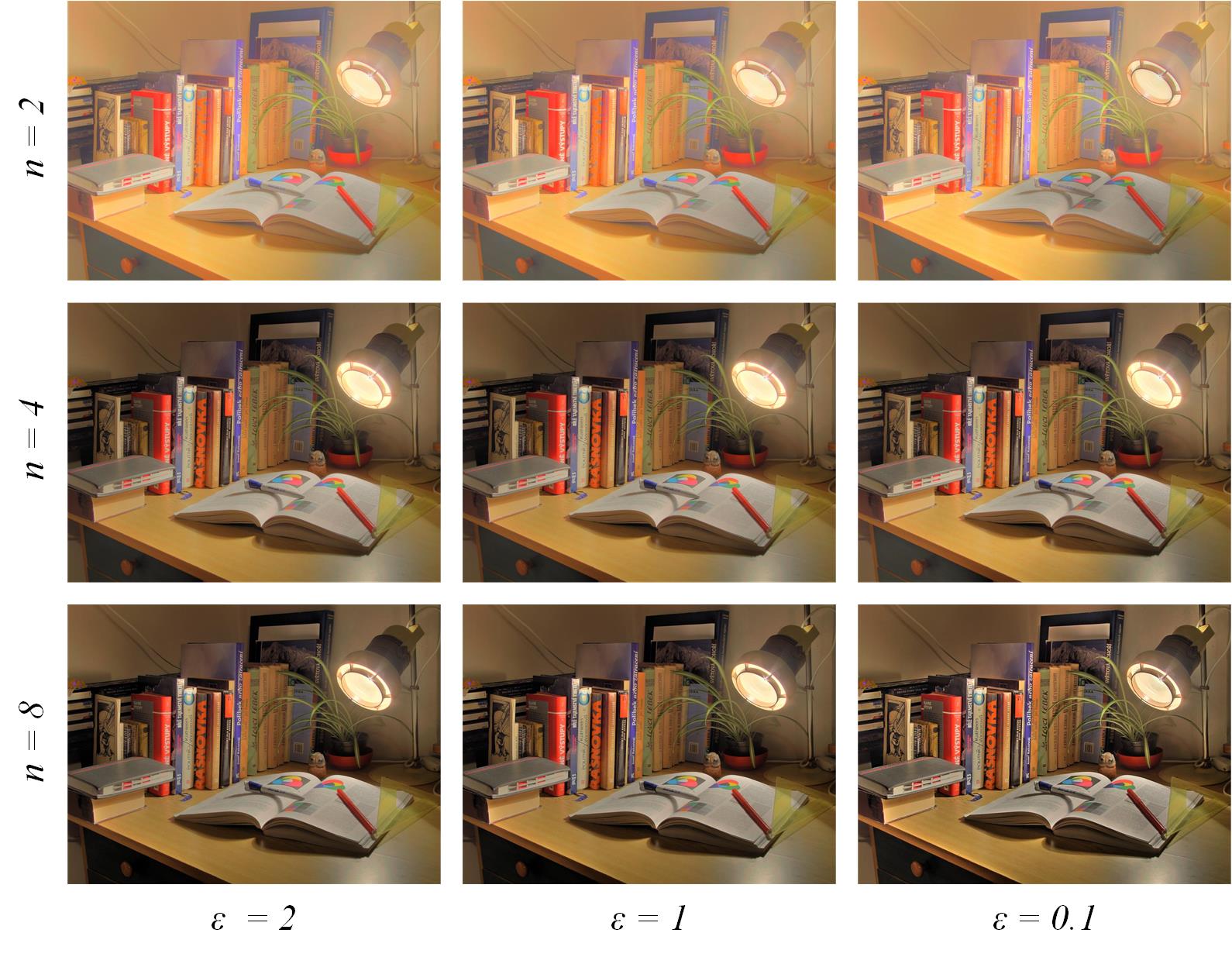}
\end{center}
\caption{Tone mapped images obtained with various $n$ and $\epsilon$ values.}
\label{ParamsChange}
\end{figure}

There are three free parameters involved in our algorithm, namely, the number of different receptive fields $s$, the number of the histogram bins $n$, and the regularization term $\epsilon$. The number of receptive fields $s$ is the most important parameter in the proposed algorithm. As we have discussed previously, to maintain the image brightness consistency, the largest receptive field should be the same as the image size. Yet to obtain possible details in different receptive fields, we need smaller receptive fields as well. We adopt the popular image pyramid method and define the relationship for any two adjacent receptive fields $w_{i+1}$ and $w_i$ as $w_{i}/w_{i+1} =2$. Consider a $2048 \times 2048$ WDR image, $s = 5$ will yield the smallest receptive field of $128 \times 128$ which is enough for the tone mapping function of Eq. 2 to preserve local details.
The effect of the number of bins $n$ and the regularization term $\epsilon$ parameters for a tone mapped image are shown in Fig. 4. Nine tone mapped images are presented in a matrix with $n$ varying vertically and $\epsilon$ varying horizontally. It is apparent that the image contrast increases as $n$ increases. The decrease of $\epsilon$ in Eq. 4 will result image with better local details. In Fig. 4, the texture of the window glass is best preserved with smallest $\epsilon $ value. We find values for $n > 5$ and $\epsilon = 0.1$ to usually produce satisfactory results, i.e. good brightness while preserving local contrast and details.

\section{Mobile-end Implementation}
The proposed algorithm is implemented in the iOS platform as an application. The application first captures four different exposures and then the four images 
are used to recover a WDR image using the algorithm proposed by Paul Debevec \cite{debevec2008recovering}. 

The generated WDR image is converted to luminance channel using the following equation:

\begin{equation}
i = 0.2126*R + 0.7152*G + 0.0722*B
\label{m-e representation}                                                                                            
\end{equation}

where $R$, $G$, and $B$ are the red, green and blue channel of the WDR image, respectively. Then, the tone mapped image is generated by applying the proposed algorithm on the luminance image $i$. At last, the color information is restored using the same color restoration function used in \cite{fattal2002gradient}:
\begin{equation}
c_{out} = (\frac{C_{in}}{L_{in}})^{sat} L_{out}
\label{m-e representation}                                                                                            
\end{equation}
where $C = R, G, B$ represents the three color channels, and $L_{in}$ , $L_{out}$ denote the luminance before and after WDR tone mapping, respectively. $sat$ is a parameter controlling color saturation that is set as $0.6$. In the following, we first briefly introduce the programming model and then will focus on the GPU implementation of the tone mapping algorithm.

\begin{figure}[tb]
\begin{center}
   \includegraphics[scale = 0.55]{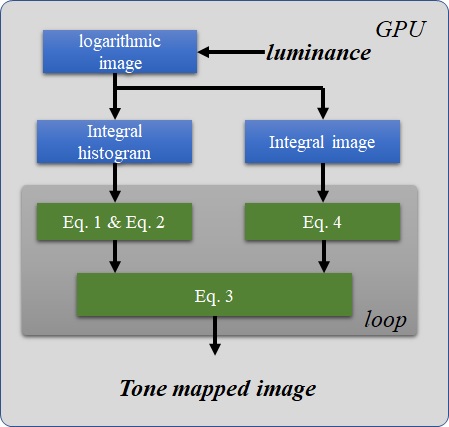}
\end{center}
   \caption{Main building blocks of GPU implementation.}
   \label{Processing_flow}
\end{figure}

\subsection{Mobile-end GPU Implementation}

The algorithm is implemented using Apple’s GPU application programming interface, Metal. Metal allows for the parallel processing of data much like OpenGL or other shader programming languages.
In Metal, kernel function is the basic function that runs in SIMD fashion in GPU. Each instance of kernel function is called a thread. Threads are further organized into \textit{threadgroups} that are executed together and can share a common block of memory. 
Since the proposed algorithm is pixel-parallel, then we can assign every pixel a thread to tone map itself and all threads share the common memory which is the calculated integral image and integral histogram.
The overall implementation flow of the algorithm is shown in Fig. \ref{Processing_flow}. The input and output of the algorithm are the calculated lumniance image and tone mapped lumniance image, respectively. The blue blocks indicate functions that execute only once during the computation. They are used to calculate the logarithmic image, integral histogram $H$ and integral image $I(i)$ and $I(i^2)$, respectively. \textit{MPSImageIntegral} function that is provided by Metal Performance Shaders is used to calculate the integral image and integral histogram. The core functions are labeled with green color in Fig. \ref{Processing_flow}.

Our implementation calculates different scales in sequence, the functions that are required in this procedure are labeled as green, and they will be called for $s$ times in a main loop (shaded gray). The pseudocode code of Algorithm 1 shows how the Eq. 4 is implemented in GPU as kernel function.  $I_{\sum}$ and $I^2_{\sum}$ are the computed integral image and of $i$ and $i^2$, respectively. They can be accessed by all threads. $grid.x$ and $grid.y$ are the vertical and horizontal coordinates of current thread. Line 2 to line 5 are used to calculate coordinates of the four corner pixels of Eq. 6 and Eq. 9. $|w|$ calculates the number of pixels in the receptive fields. Eq. 6 and Eq. 9 are executed in line 13 and line 18. Line 19 implement Eq. 10 and Eq. 4 is carried out in line 20. The implementation of algorithm 1 only employs basic addition, subtraction and indexing operations of GPU.

\begin{algorithm}[tb]  
  \caption{Eq. 4 GPU implementation pseudocode}  
  \begin{algorithmic}[1]  
    \Require  
      integral of luminance image $I_{\sum}$;  
      integral of luminance image $I^2_{\sum}$;  
      regularization term $\epsilon$;
      current scale $j$;
    \Ensure  
     weight, $W_{w_{j}}$;    
    \For{all pixel in $L_{\sum}$ and $L^2_{\sum}$} 
    \State $x_{-} = grid.x - (s_j.x + 1)$
    \State $x_{+} = grid.x + s_j.x$
    \State $y_{-} = grid.y - (s_j.y + 1)$
    \State $y_{+} = grid.y + s_j.y$
   
    
    \State $x_{in} = 2s_j.x+1$
    \State $y_{in} = 2s_j.y+1$
    
	\State $|w| = x_{in} * y_{in}$    
	
    \State $A \leftarrow  I_{\sum(i,j)}(x_{-},y_{-}) $
    \State $B \leftarrow  I_{\sum(i,j)}(x_{+},y_{-}) $
    \State $C \leftarrow  I_{\sum(i,j)}(x_{+},y_{+}) $
    \State $D \leftarrow  I_{\sum(i,j)}(x_{-},y_{+}) $
    \State $\mu_{1} = A+C-B-D  $

    \State $A \leftarrow I^2_{\sum(i,j)}(x_{-},Y_{-}) $
    \State $B \leftarrow I^2_{\sum(i,j)}(x_{+},Y_{-}) $
    \State $C \leftarrow I^2_{\sum(i,j)}(x_{+},Y_{+}) $
    \State $D \leftarrow I^2_{\sum(i,j)}(x_{-},Y_{+}) $
    \State $\mu_{2} = A+C-B-D $
    \State $\sigma_w = (\mu_2 - \mu^2_1)/|w| $

    \State $W_{w_{j}} = {\sigma_w}/({\sigma_w + \epsilon})$
    \EndFor 
    \label{code:HIST} \\  
    \Return $W_{w_{j}}$;  
  \end{algorithmic}  
\end{algorithm}

Eq. 1, Eq. 2 and Eq. 3 are also implemented as kernel functions because they only depend on the pixel value and the integral histogram.
The parallel processing feature of the algorithm greatly reduce the computation complexity and cost of the implementation. Evaluation results of the GPU implementation will be shown in the following section.

\section{Experimental Results}
In this section, we present image quality comparison of our algorithm as well as the performance of the GPU implementation.

Our algorithm is compared with three tone mapping algorithms --- local Laplacian-based tone mapping by Paris \textit{et al.} \cite{paris2015local}, the fast bilateral filtering tone mapping method by Durand \textit{et al.} \cite{durand2002fast} and edge-preserving multi-scale decomposition algorithm proposed by Gu \textit{et al.} \cite{gu2013local}. The three algorithms reported state-of-the-art tone mapped image quality. We use the codes of \cite{paris2015local} and \cite{durand2002fast} provided by the authors themselves. For Gu's method, we implemented the algorithm based on the code of \cite{he2010guided}. In the image quality assessment, all compared algorithms use their default parameter settings. Our algorithm use parameter setting $n = 5$, $s= 5$ and $\epsilon = 0.1$. For application performance evaluation, we tested our application on an iPhone 6 plus device. 

\begin{table}
\centering
\vfill
\caption{\textsc{TMQI naturalness Scores}}
\begin{tabular}{ l  c  c  c  c}
\hline
\hline
Image & \tabincell{c}{ Durand \textit{et al.} \\ \cite{durand2002fast}} & \tabincell{l}{ Gu \textit{et al.} \\ \cite{gu2013local}} & \tabincell{c}{ Paris \textit{et al.} \\  \cite{paris2015local}} & Proposed \\
\hline
BristolBridge  	&0.0802 &0.3262 	&0.2160 	&\textbf{0.8558}	\\ 
ClockBuilding 	&0.8032	&\textbf{0.8829}		&0.6845		&0.8345	\\
CrowFootGlacier &0.0887	&0.4725 	&0.5205   	&\textbf{0.7416}	\\ 
DomeBuilding	&0.4134	&0.9108		&0.4570 	&\textbf{0.9799}   	\\ 
FribourgGate	&0.3923	&\textbf{0.86837}	&0.8239		&0.8541		\\
MontrealStore	&0.3959	&\textbf{0.9725}	&0.4489		&0.9383		\\
Moraine2		&0.1260	&\textbf{0.7519}		&0.2984		&0.6234	\\
Oaks			&0.2199	&0.9281		&0.4045	&\textbf{0.9770}		\\
StreetLamp		&0.5590	&0.7952		&0.4141		&\textbf{0.8145}	\\
Vernicular		&0.4190	&0.6256		&\textbf{0.7458}	&0.6023 	\\
\hline 
Average			&0.3498	&0.7534		&0.5014		&\textbf{0.8221} 	\\
\hline 
\hline
\end{tabular}
\label{N score}
\end{table}

\begin{table}
\centering
\vfill
\caption{\textsc{TMQI Structural Similarity Scores}}
\begin{tabular}{ l  c  c  c  c}
\hline
\hline
Image & \tabincell{c}{Durand \textit{et al.} \\ \cite{durand2002fast}} & \tabincell{l}{Gu \textit{et al.} \\ \cite{gu2013local}}& \tabincell{c}{ Paris \textit{et al.} \\  \cite{paris2015local}}& Proposed \\ 
\hline
BristolBridge  	&0.8368 &0.7947 	&0.8387 	&\textbf{0.9095}	\\ 
ClockBuilding 	&0.8780	&0.8101		&0.8795		&\textbf{0.8942}	\\
CrowFootGlacier &0.9274	&0.8181 	&0.9422   	&\textbf{0.9424}	\\ 
DomeBuilding	&0.7364	&0.6968		&0.7498		&\textbf{0.7933}   	\\ 
FribourgGate	&0.9362	&0.9045		&0.9419		&\textbf{0.9448}	\\
MontrealStore	&\textbf{0.9330}	&0.8940		&0.9321		&0.9213	\\
Moraine2		&0.9144	&0.9014		&0.9242		&\textbf{0.9576}	\\
Oaks			&0.9333	&0.8832		&0.9578		&\textbf{0.9703}	\\
StreetLamp		&0.8844	&0.8694		&0.8654		&\textbf{0.9436}	\\
Vernicular		&0.9117	&0.8946		&0.9255		&\textbf{0.9360} 	\\
\hline 
Average			&0.8892	&0.8467		&0.8958		&\textbf{0.9213}	\\
 \hline 
 \hline
\end{tabular}
\label{psychophysical_experimentation}
\end{table}

\subsection{Image Quality Assessment}
It can be understood from the earlier explanation that our algorithm utilizes all available display levels in each local processing. This significantly increases image brightness, especially in dark regions. Our experiments tested different WDR images to evaluate the image quality of the tone mapped images. One example is shown in Fig. 6. It shows one image tone mapped with different algorithms. In reading order, the images are tone mapped with Durand \textit{et al.}, Gu \textit{et al.}, Paris \textit{et al.}, and the proposed algorithm, respectively. A visual impression can tell that our algorithm gives an image with higher brightness value over the other three images. In fact the average brightness value for the four images is 93.78, 120.92, 98.01 and 122.14 respectively. Despite the fact that the image generated with Gu \textit{et al.} has similar overall brightness, it lacks global contrast and many unwanted details are enhanced which makes the image look unnatural. 

For objective assessment, we use the tone-mapped image quality index (TMQI) \cite{yeganeh2013objective} to calculate an overall quality score that combines a multi-scale structural fidelity measure and a measure of image naturalness. The structural fidelity measure is a full-reference assessment based on the structural similarity (SSIM) index. The naturalness measure is a no-reference assessment based on statistics of good-quality natural images. 
The results of the naturalness score, structural fidelity score,  and the overall score are listed in Table I, Table II and Table III. The winner algorithm's score is shown in bold font. In the naturalness score, our algorithm scores highest in 5 images and achieves an average value of 0.8221 for which is highest among the tested algorithms. In terms of structural similarity, our algorithm wins 9 out of 10 images and achieves an average score of 0.9213. In terms of overall quality, our algorithm produces the best scores for 9 images.

\begin{table}
\centering
\vfill
\caption{\textsc{TMQI Overall Scores}}
\begin{tabular}{ l  c  c  c  c}
\hline
\hline
Image & \tabincell{c}{Durand \textit{et al.}\\ \cite{durand2002fast}}& \tabincell{l}{Gu \textit{et al.} \\ \cite{gu2013local}}& \tabincell{c}{ Paris \textit{et al.} \\  \cite{paris2015local}}& Proposed \\ 
\hline
BristolBridge  	&0.7921 &0.8369 	&0.8265 	&\textbf{0.9564}	\\ 
ClockBuilding 	&0.9403	&0.9334		&0.9224		&\textbf{0.9493}	\\
CrowFootGlacier &0.8187	&0.8705 	&0.9120   	&\textbf{0.9477}	\\ 
DomeBuilding	&0.8362	&0.9038		&0.8480		&\textbf{0.9426}   	\\ 
FribourgGate	&0.8877	&0.9570		&0.9600		&\textbf{0.9652}	\\
MontrealStore	&0.8875	&0.9692		&0.8969		&\textbf{0.9715}	\\
Moraine2		&0.8254	&0.9387		&0.8666		&\textbf{0.9329}	\\
Oaks			&0.8525	&0.9600		&0.8954		&\textbf{0.9894}	\\
StreetLamp		&0.9034	&0.9368		&0.8731		&\textbf{0.9590}	\\
Vernicular		&0.8863	&0.9171		&\textbf{0.9440}	&0.9240 	\\
\hline 
Average			&0.8630	&0.9223		&0.8945		&\textbf{0.9538}	\\
 \hline 
 \hline
\end{tabular}
\label{psychophysical_experimentation}
\end{table}

\begin{figure}[h]
\begin{center}
   \includegraphics[scale = 0.8]{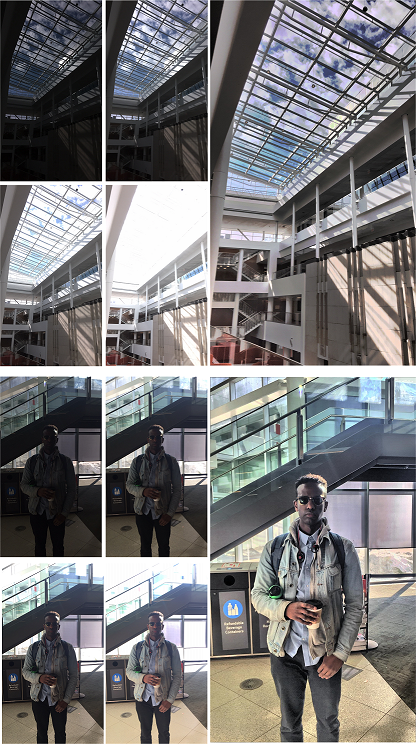}
 \end{center}
   \caption{Field test results. The left two columns show four different exposure images, the right image shows the tone mapped image with proposed algorithm. }
   \label{}
\end{figure}

Field assessment of the image quality has also been conducted using the developed iOS application. Fig. 6 shows two sets of examples. The first scene is taken in a commonly seen extreme light condition. The four images with different exposure are shown on the two left columns and the tone mapped image is shown on the right-most column. Our algorithm produces a bright image with clear details --- For example, the outdoor cloud and the frame of skylight window are clearly visible. The inner structure of the building such as stairs and railing are also clearly shown in the tone mapped image. Another scene is a portrait image because taking portraits is the most common use of phone camera. The bottom image of Fig. 6 shows an example image of a portrait taken by our application. This image is taken under backlight condition, it is a challenging situation because the strong background light could make the portrait dark or even invisible. The tone mapped image in Fig. 6 shows the portrait clearly and the detail in the background is also well preserved. This experiment indicates that our application is very suitable for imaging under extreme light conditions.

\subsection{Application Performance Assessment}

\begin{table}
    \centering
    \vfill
    \captionsetup{justification=centering}
    \caption{\textsc{Measured Processing time with different Settings}}
    \begin{tabular}{|c|c|c|c|c|} 
    \hline 
    \hline
    & Number of scales & $480 \times 640$  & $720 \times 1280$  &  $1080 \times 1920$  \\ 
    \hline
\multirow{3}{*}{$n =3 $}   & $s = 3$ & 122.11  & 303.79  &  637.90 \\ 
\cline{2-5}
                  & $s = 4$  & 145.82  & 351.13  & 788.27 \\ 
\cline{2-5}
                  & $s = 5$  & 166.91 & 402.36 &  865.69 \\ 
\hline \hline
\multirow{3}{*}{$n = 4$} & $s = 3$ & 130.33  & 284.74  & 677.69  \\ 
\cline{2-5}
                  & $s = 4$ & 174.73 & 359.80 &  793.60 \\ 
\cline{2-5}
                  & $s = 5$ & 175.14 & 422.65 &  920.95 \\ 
\hline \hline
\multirow{3}{*}{$n = 5$} & $s = 3$ & 139.69 & 293.37  &  869.33  \\ 
\cline{2-5}
                  & $s = 4$ & 187.64 & 355.05 &  886.96 \\ 
\cline{2-5}
                  & $s = 5$ &212.44  &432.90  &  1016.26 \\
\hline \hline
\end{tabular}
\label{}
\end{table}

Ideally, we hope to compare the application performance with some other works which are implemented also in mobile-end. However, most tone mapping algorithms reported are implemented on desktop platforms. It would be unfair to compare implementations on different platforms. Hence, we focus our analysis on the mobile-end implementation.

The proposed algorithm has three parameters, namely the number of scales $s$, the number of bins $n$ and the regularization term $\epsilon$. A detailed performance analysis should fully consider the variation of the three parameters. The regularization term $\epsilon$ is a scalar value which does not affect neither the memory requirement nor the total amount of computation, so it will not affect the performance of our application. However, the number of scales affects the computation burden and the number of bins $n$ affects the required memory usage. To evaluate the two parameters $s$ and $n$ 's effect on the processing speed, we tested $9$ parameter sets, varying $s$ from 3 to 5 and $n$ from 3 to 5. The evaluation results of three commonly used resolutions $480\times 640$, $720\times 1280$ and $1920 \times 1080$ are shown in Table IV. For identical parameter settings, the processing time increases mostly linearly as the image resolution increases. 
For example, the processing times are about $200$ \textit{ms}, $400$ \textit{ms} and $1000$ \textit{ms} for the three different resolutions when $n = 5, s = 5$. The influence of the $n$ parameter on the processing time is limited. Under same image resolution with a fixed number of scales, the processing time only fluctuates in a very limited range. 
In $720\times 1280$ resolution, the processing time are $402$ \textit{ms}, $422$ \textit{ms} and $432$ \textit{ms} when $s$ equals to $5$ and $n$ equals to 3, 4 and 5, respectively. 
In $1080 \times 1920$ resolution, the processing time is $788$ \textit{ms}, $793$ \textit{ms} and $886$ \textit{ms} when $s$ equals to $4$ and $n$ equals to $3$, $4$ and $5$, respectively. 
From Table IV, we can see that the scale parameter $s$ has greater influence on the application processing time than the $n$ parameter.
In $480\times 640$ resolution, the application processing time is $122$, $145$ and $166$ \textit{ms} when the $n$ is equal to 3 an $s$ is equal to 3, 4, 5, respectively. 
As for the $720\times 1280$ and $1080 \times 1920$ resolution, the processing time can increase about $50\%$ when scale $s$ changes from 3 to 5. It is easy to conclude that the greatest factor that affects the application processing time is the image resolution. Parameter $s$ affects the application time secondarily while parameter $n$ has only very little influence. 

\begin{figure}[tbh]
\begin{center}
   \includegraphics[scale = 0.4]{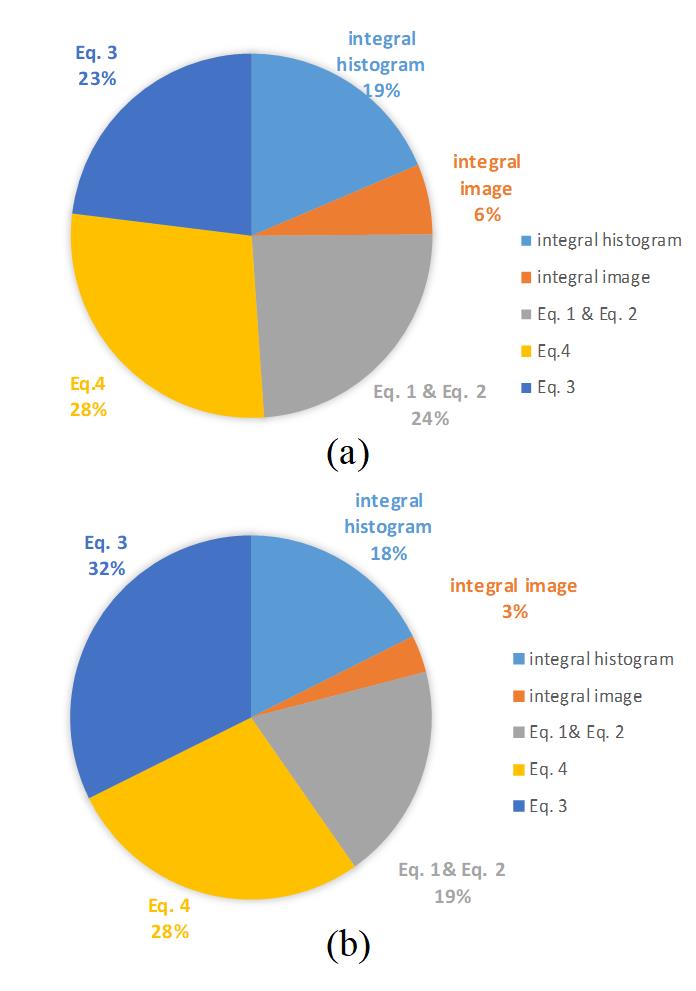}
\end{center}
   \caption{Computational breakdown of the application.}
   \label{Fig_9}
\end{figure}

A processing time breakdown using two different parameter settings is shown in Fig. 7. Fig. 7 (a) is the processing time breakdown when $n = 5$, $s = 5$ and image resolution is $1920 \times 1080$. Fig. 7 (b) is the processing time breakdown when $n = 3$, $s=3$ and image resolution equals to $480\times 640$. 
We choose the two extreme cases to show the rough proportion of each function. Since Eq. 3, Eq. 4 and Eq. 1 \& Eq. 2 will be looped for several times, hence, they consumes the majority of processing time.
They account for 23\%, 28\% and 24\% of the total processing time in Fig. 7 (a) and 32\%, 28\% and 19\% in Fig. 7(b). The \textit{integral histogram} and \textit{integral images} only need to be computed once in our algorithm and they consume 19\% and 6\% of processing time in Fig. 7 (a), 18\% 3\% in Fig. 7(b) respectively. 

\section{Conclusion}
A tone-mapping algorithm based on integral image and integral histogram for tone mapping of wide dynamic range images is presented. The algorithm is motivated by the local processing feature of the human visual system. It adopts multiple receptive fields to combines global image consistency and local image details into one final image.
Quality evaluation as well as field testing were carried out and discussed in detail. In the objective assessment, results showed that the proposed algorithm performed best in both structural similarity score and naturalness score. Hence, highest TMQI index scores were achieved by our algorithm compared to the three other state of the art algorithms. 
In the application field test, our algorithm also produced appealing image which displayed scene details. Mobile GPU implementation of the proposed algorithm was presented, which can perform tone mapping of typical 1080P WDR color images at about 1 second, thus making it suitable for mobile phone users. 

\section*{Acknowledgments}
The authors would like to thank the Alberta
Innovates Technology Futures (AITF) and Natural Sciences and
Engineering Research Council of Canada (NSERC) for supporting this research.

{
\footnotesize
\bibliographystyle{IEEEtranTIE}
\bibliography{BIB_1x-TIE-2xxx}
}

\end{document}